\documentclass[10pt]{iopart}

\usepackage{iopams}
\usepackage{hyperref}
\hypersetup{colorlinks=true, urlcolor=red, citecolor=blue, linkcolor=blue}
\usepackage{graphicx}
\usepackage{subfigure}
\usepackage{epstopdf}
\usepackage{epsfig}
\usepackage{cite}
\usepackage[monochrome]{color}

\begin{document}

\renewcommand*{\DefineNamedColor}[4]{%
   \textcolor[named]{#2}{\rule{7mm}{7mm}}\quad
  \texttt{#2}\strut\\}

\definecolor{red}{rgb}{1,0,0}
\definecolor{cyan}{cmyk}{1,0,0,0}

\title[Density Matrix Recursion Method]{Density Matrix Recursion Method: Genuine multisite entanglement distinguishes odd from even quantum spin ladder states}

\author{Himadri Shekhar Dhar\(^{1}\), Aditi Sen(De)\(^{2}\), and Ujjwal Sen\(^{2}\)}

\address{\(^1\)School of Physical Sciences, Jawaharlal Nehru University, New Delhi 110 067, India\\
\(^2\)Harish-Chandra Research Institute, Chhatnag Road, Jhunsi, Allahabad 211 019, India
}

\ead{ujjwal@hri.res.in}

\date{05 October 2012}

\begin{abstract}
%
%

We introduce an analytical iterative method, the density matrix recursion method, to generate 
arbitrary reduced density matrices
of
superpositions of short-ranged dimer coverings on periodic \textcolor{red}{or non-periodic quantum} spin-1/2 ladder lattices, with 
an arbitrary number of legs. \textcolor{red}{The method can be used for 
calculating bipartite as well as multipartite physical properties, 
including bipartite and multipartite entanglement.}
We apply this technique to distinguish between even- and odd-legged ladders. Specifically, we show that while genuine 
multipartite entanglement decreases with increasing system-size for the even-legged ladder states, it does the opposite for odd-legged ones.
\end{abstract}


\submitto{\NJP}

\maketitle 

\section{INTRODUCTION}
The quantum spin ladder is an interesting platform to investigate quantum many body systems in the intermediate sector between the one- and two-dimensional lattice 
structures \cite{dag}. 
The possibility of relating doped even-legged quantum spin ladders to high-temperature superconductivity \cite{super,bhaskar,expt} make such quantum systems extremely 
important.
The characteristic pseudo 2-D structure of ladders evokes considerable interest in several other areas of condensed matter physics \cite{Scal,white} and 
quantum information  \cite{QI}. 
On the other hand, 
concepts like entanglement \cite{Hor} and other quantum correlations \cite{kav} have been applied to understand quantum critical phenomena in spin 
systems \cite{Faz}, including in  
spin ladders 
\cite{QI}. Properties of such systems have also been tested experimentally in several systems or proposals thereof have been presented, 
including in compounds and  
cold gas \cite{ladder-expt}. \textcolor{red}{Ladder 
states formed by the superposition of short-range dimer coverings, also known as short-range resonating valence bond (RVB) states, can be simulated by using atoms in
 optical lattices \cite{OL}, and, as shown recently, by using interacting photons \cite{WP}. These short-range RVB ladder states are believed to be possible ground states 
of \textcolor{red}{certain} undoped Heisenberg spin ladders, supported by various 
methods which include mean field theory \cite{mf}, quantum Monte Carlo \cite{expt, qmc}, and Lanczos \cite{Scal, lan}.} 

A striking  feature of the quantum spin ladder is that the interpolation from the 1-D spin chain to the 2-D square lattice by gradually increasing the number of legs is 
not straighforward. \textcolor{red}{For example,} the quantum characteristics of the Heisenberg ladder ensures that the odd- and the even-legged ladder ground states
have different correlation 
properties. ``Even ladders'' have a finite-gapped ground state excitation and exponential decay of two-site correlations, while ``odd ladders'' are gapless and have 
power-law decay \cite{oldpaper,white}. Such effects also occur in quantum spin liquids, such as discussed in  \cite{Referee-report}. 
The results of Heisenberg ladders cannot therefore be directly extrapolated to the 2-D regime. This difference in quantum 
characteristic of odd- and even-legged ladders may lead to interesting features in the entanglement properties of the system. However, calculating the bipartite or 
multipartite entanglement in large-sized multi-legged quantum spin ladder states formed by the superposition of short-range dimer coverings is numerically challenging. 
Within a dimer-covering approach \cite{white}, it is possible to derive energy density and spin-correlation functions
for two-legged ladders by using generating functions \cite{Fan} or state iterations \cite{del}.
Approximate 
solutions may also be obtained for the four-legged ladder \cite{del2}. 
For further work in this direction, including density matrix renormalization group calculations in multi-legged Heisenberg ladders, see 
\cite{patraghat}.

In this paper, we introduce an analytical iterative technique, the 
``density matrix recursion method'' (DMRM), \textcolor{red}{to obtain the reduced density matrix of an arbitrary number of sites of a state
 on a quantum spin-1/2 ladder with an arbitrary number of legs and with both open and periodic boundary conditions. We consider the spin-1/2 ladder
 state to be a superposition of short-ranged dimer coverings.}
Specifically, 
within a dimer covering approach for the state of the quantum spin ladders,
we find separate iterative formalisms for even and odd ladders. 
These partial density matrices can be used to calculate and study the scaling and behavior of single-, two-, and multi-site physical properties of the whole
 spin system, including 
two-site correlations and bipartite as well as multipartite entanglement. 
\textcolor{red}{The iterative method introduced here can also be a potential tool for studying properties of reduced density matrices of general large superposed 
multipartite quantum states and hence can be used as an efficient method for investigating quantum correlations of multipartite quantum states.}

We then apply our method to obtain the nature of  genuine multipartite entanglement, quantified by 
the 
\emph{generalized geometric measure} (GGM) \cite{GGM}. 
\textcolor{blue}{An understanding of the multipartite entanglement content 
is potentially advantageous, in assessing the importance of the corresponding state for future applications,
over that of 
bipartite  measures,  as the former \textcolor{red}{takes into account the} distribution of information between the multisite sub-regions 
\textcolor{red}{of the entire system} and the 
hidden correlations are not ignored due to tracing out \cite{qwertyu}.}  
We compare the odd- and even-legged ladder states by using genuine multisite entanglement, and find that the GGM of 
odd-legged ladder states increases with system-size while it decreases in the even ones. 
%
%
\textcolor{blue}{We also find that the convergence of the GGM, for a given odd- or even-legged ladder, occurs for  relatively small ladder lengths of the 
corresponding ladder.}

The paper is organized as follows. We introduce the model state in Sec. \ref{aaj-chicken-curry}. The genuine multiparty entanglement measure is 
defined in Sec. \ref{sec-ggm}. The main results are presented in sections \ref{sec-dmrm}, \ref{odd}, and \ref{sec-evo}. In particular, the density matrix recursion methods 
for 
even and odd ladders are respectively introduced in sections \ref{sec-dmrm} and \ref{odd}. We present a conclusion in Sec. \ref{sec-conclu}.

\section{Model State}
\label{aaj-chicken-curry}
Within a short-range dimer-covering approach, the state of the quantum spin-1/2 ladder consists of an equal superposition of nearest neighbor directed dimer pairs on the spin-1/2 lattice, also called the resonating valence bond (RVB) state. 
The ladder lattice under consideration can be divided into sublattices (see Fig. 1), A and B, in such a way that all 
 nearest neighbor (NN) sites of sublattice A belong to sublattice B, and vice versa. Such a lattice is called a \emph{bipartite lattice}.   
Each bipartite lattice site is occupied by a spin-1/2 qubit.
The multi-legged quantum spin-1/2 ladder consists of coupled parallel spin chains with $\mathcal{N}$ sites on each chain labelled from $1$ to $\mathcal{N}$. 
The number of chains, referred as legs, of the ladder are labelled as 1 to $M$. The total number of spin-1/2 sites in the entire bipartite lattice is $s$ ($s=M\mathcal{N}$).
 The vertical chains, referred as rungs, are labelled from $1$ to $n$, similar to the sites on the chain, and each rung contains $M$ spins. 
See Fig. 1.
The periodic ladder is 
conditionally defined by allowing dimer states to be formed between rungs $1$ and $n$ \cite{Fan}.  Of course as numbers, $n=\mathcal{N}$.
The (unnormalized) quantum state under study is therefore
\cite{anders} 
\[
|\mathcal{N}\rangle=\sum_k \left[\prod_{i,j}|(a_i,b_j)\rangle\right]_k,
\] 
where $|(a_i,b_j)\rangle $
 refers to a dimer between sites $a_i$ and $b_j$
on 
the bipartite lattice. $(i,j)$ are NN sites with $i \neq j$. A product of all such dimers on the spin-1/2 lattice constitutes an RVB \emph{covering}. 
The summation refers to the superposition of all such dimer product states that give us the superposed short-range dimer covering states. 
\textcolor{red}{It is believed that these short-range dimer covered states are possible solutions of the two-dimensional antiferromagnetic Heisenberg systems based 
on the understanding of the RVB theory and possible 
explanation of the results obtained by using numerical methods like the density matrix renormalization group \cite{del,DM}.} 


\begin{figure}[htb]
\label{fig:1}
\begin{center}
\epsfig{figure= 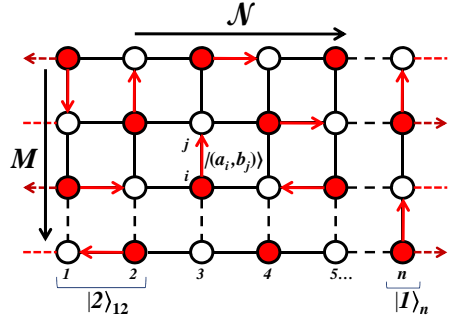,
height=.3\textheight, width=0.35\textwidth,angle=-90}
\caption{\textcolor{red}{The $M$-legged, $\mathcal{N}$-runged quantum spin-1/2 ladder in the form of a \textit{bipartite lattice}}. The solid circles are of 
sublattice $A$, while the 
hollow ones are of  $B$. The red arrows (solid) show the nearest-neighbor dimer states ($|(a_i,b_j)\rangle$) from a site in sublattice $A$ to another in $B$.
 The arrows (dashed) at the boundary indicate that the lattice is with periodic boundary condition. The dashed lines indicate that the number of legs and rungs can be extended beyond the illustrated number of sites.
The figure also shows the two-site ladder state, $|2\rangle$, and the one-site rung state, $|1\rangle$, which will be profusely used in the iteration method 
presented.}
\end{center}
\end{figure}

\section{Genuine Multipartite Entanglement and Generalized Geometric Measure}
\label{sec-ggm}
\textcolor{red}{
Quantification of multipartite entanglement plays a crucial role in quantum information processing. In the case of 
bipartite pure states, entanglement can be uniquely quantified as the von Neumann entropy of local density matrices. 
In comparison, the quantification of multipartite entanglement involves characterizations and criteria of multipartite pure quantum states and cannot
 be uniquely defined \cite{Hor}. For pure states, a genuinely multipartite entangled state can be defined as a multiparty pure quantum state that cannot be
 expressed as a product across any bipartition. The Greenberger-Horne-Zeilinger \cite{GHZ} and the W \cite{W} states are the common examples of genuine multipartite
 entangled states. Genuine multipartite entanglement is an important resource from the perspective of many-party quantum communication \cite{physnews} and is a potential
 resource for large-scale quantum computation \cite{eck}. Cluster states are multipartite entangled states that form promising candidates in building a one-way quantum 
computer \cite{com}. 
Multipartite entanglement also has fundamental applications in understanding the role of entanglement in many-body physics \cite{Faz}.}

\textcolor{red}{
As mentioned above, a genuine multisite entangled pure quantum state is one which is entangled across every partition,  into two subsets,
 of the set of the observers involved. A natural definition of the amount of genuine multipartite entanglement of a pure multiparty quantum state is therefore the
 minimum distance of that state from the set of all states that are not genuinely multipartite entangled. A widely used measure of distance is the fidelity subtracted 
from unity \cite{eisob-distance}. The generalized geometric measure (GGM) of an \(N\)-party pure quantum state $|\phi_N\rangle$ is defined as the minimum of
 this ``fidelity-based distance" of $|\phi_N\rangle$ from the set of all states that are not genuinely multiparty entangled \cite{GGM,amadernext}. In other words,
 the GGM of $|\phi_N\rangle$, denoted by $\mathcal{E}( |\phi_N\rangle )$, is given by
\begin{equation}
\mathcal{E} ( |\phi_N\rangle ) = 1 - \Lambda ^ 2_{\max} (|\phi_N\rangle ),
\label{GM}
\end{equation}
where  \(\Lambda_{\max} (|\phi_N\rangle ) = \max | \langle \chi|\phi_N\rangle |\), with  the maximization being over all \(N\)-party quantum states 
\(|\chi\rangle\) that are not genuinely multipartite entangled. 
}

It was shown in Ref. \cite{GGM, amadernext} that the GGM of a multiparty pure state can be effectively calculated by using
\begin{equation}
{\mathcal E} (|\phi_N \rangle ) =  1 - \max \{\lambda^2_{ A: B} |  A \cup  B = \{1,2,\ldots, N\},  A \cap  B = \emptyset\},
\end{equation}
where \(\lambda_{A:B}\) is  the maximal Schmidt coefficient in the bipartite split \(A: B\) of \(|\phi_N \rangle\). 
The GGM can therefore be calculated once the reduced density matrices of the pure quantum state is obtained. We use the GGM to quantify the amount of genuine multisite entanglement present in multi-legged ladder states of different sizes.

\section{DMRM: The even ladder}
\label{sec-dmrm}

We will now describe the iterative methods for generating arbitrary local density matrices
for the even and the odd ladders.
We begin with the even ladders.
%
The recursion relation for the non-periodic (i.e. with open boundary conditions) even-legged ladder can be written as
\begin{eqnarray}
\label{nonper}
|\mathcal{N}+2 \rangle &=& |\mathcal{N}+1\rangle |1\rangle_{n+2} + |\mathcal{N}\rangle |\bar{2}\rangle_{n+1,n+2}\nonumber\\
&=&|\mathcal{N}\rangle |2\rangle_{n+1,n+2} + |\mathcal{N}-1\rangle|\bar{2}\rangle_{n,n+1}|1\rangle_{n+2},
\end{eqnarray}
where $|\mathcal{N}\rangle$ is an even-legged ladder containing $\mathcal{N}$ sites in each chain and corresponds 
to the rungs numbered from 1 to \(n\). On the other hand, e.g. \(|\mathcal{N}\rangle_{2,n+1}\) represents an even-legged ladder containing 
\(\mathcal{N}\) sites in each chain, and corresponds to the rungs numbered from 2 to \(n+1\).
$|1 \rangle$ is a vertical chain containing $M$ spins representing a single rung of the ladder with the dimer coverings between adjacent sites. 
$|2\rangle_{n+1,n+2}$ is a two-rung even-legged ladder of rungs $n+1\) and \(n+2$. Also, 
\[
|\bar{2}\rangle_{n+1,n+2} =|2\rangle_{n+1,n+2} - 
|1\rangle _{n+1}|1\rangle_{n+2}. 
\]
\textcolor{red}{The term $|\bar{2}\rangle_{n+1,n+2}$ contains all the coverings of a two-rung even-leg ladder $|2\rangle_{n+1,n+2}$, apart from the coverings formed by the product of the two rungs $|1\rangle _{n+1}|1\rangle_{n+2}$. The subtraction removes possible repetition of states between the two terms in the recursion for $|\mathcal{N}+2 \rangle$ in relation (\ref{nonper}).}
\textcolor{red}{We further note that }if no site numbers are mentioned in subscript of $|\mathcal{N}\rangle$, it is to be considered to be from $1$ to $n=\mathcal{N}$.

The periodic boundary condition entails that the ladder also form dimer states between the rungs $1$ and $n$. 
Accounting for the additional states in the ladder system, due to the periodicity,
 the recursion relation can be modified into
\begin{eqnarray}
\label{eperiodic}
|\mathcal{N}+2 \rangle_P = |\mathcal{N}+2\rangle_{1,n+2} + |\mathcal{N}\rangle_{2,n+1}|\bar{2}\rangle_{n+2,1},
\end{eqnarray}
where the subscript $P$ stands for a periodic ladder state. Throughout the paper,
 a state without a  subscript \(P\) will imply a non-periodic state. The density matrix for the periodic ladder state can be calculated by using the recusrsion 
relation 
in Eq. (\ref{eperiodic}):
\begin{eqnarray}
\rho^{(\mathcal{N}+2)}_P &=& |\mathcal{N}+2\rangle \langle \mathcal{N}+2|_P \nonumber\\
&=& |\mathcal{N}+2\rangle \langle \mathcal{N}+2|+|\mathcal{N}+2 \rangle \langle \mathcal{N}|_{2,n+1} \langle \bar{2}|_{n+2,1}\nonumber\\
&+&|\mathcal{N}\rangle_{2,n+1}|\bar{2}\rangle_{1,n+2} \langle \mathcal{N}+2|\nonumber\\
&+&|\mathcal{N}\rangle\langle \mathcal{N}|_{(2,n+1)}|\bar{2}\rangle \langle \bar{2}|_{(1,n+2)}.
\end{eqnarray}
If, for example, we trace out all but two rungs (say $n+1$ and $n+2$) from the state, we will obtain a two-rung (mixed state) ladder containing $2M$ spins, 
where $M$ is the number of legs  of the ladder. Note that \(M\) is even here. 
These reduced states can then be used to calculate the multisite entanglement properties of the ladder system. 
In particular, 
\begin{equation}
\rho^{(2)}_{P(n+1,n+2)} = \tr_{1..n}(|\mathcal{N}+2\rangle \langle \mathcal{N}+2|_P)
\end{equation}
 is the two-rung state of the periodic ladder. 
Now, using relation (\ref{nonper}), we can obtain the trace for the non-periodic part to have
\begin{eqnarray}
  \tr_{1..n}(|\mathcal{N}+2\rangle \langle \mathcal{N}+2|) = \tr_{1..n} \large[|\mathcal{N}\rangle \langle \mathcal{N}||2 \rangle \langle 2|_{(n+1,n+2)} \nonumber \\
 + |\mathcal{N}-1\rangle \langle \mathcal{N}-1| |\bar{2}\rangle \langle \bar{2}|_{(n,n+1)}|1 \rangle \langle 1|_{(n+2)}  \nonumber \\
 + |\mathcal{N}\rangle |2\rangle_{n+1,n+2} \langle \mathcal{N}-1| \langle\bar{2}|_{n,n+1} \langle 1|_{n+2}  \nonumber \\ 
 + |\mathcal{N}-1\rangle|\bar{2}\rangle_{n,n+1}|1\rangle_{n+2}\langle\mathcal{N}|\langle 2|_{n+1,n+2}\large].    \,\,\,\,
\end{eqnarray}

Tracing over the rungs 1 to \(n\), we get
\begin{eqnarray}
\label{np}
\rho^{(2)}_{(n+1,n+2)}&=& \mathcal{Z}_{\mathcal{N}} |2\rangle \langle 2|_{(n+1,n+2)} + 
\mathcal{Z}_{\mathcal{N}-1} \bar{\rho}_{n+1} \otimes |1\rangle\langle 1|_{(n+2)} \nonumber\\
&+& (|2 \rangle_{n+1,n+2} \langle 1|_{n+2} \langle \xi_{\mathcal{N}} |_{n+1} + h.c.),
\end{eqnarray}
where $\mathcal{Z}_{\mathcal{N}}= \langle\mathcal{N}|\mathcal{N}\rangle$, 
$\bar{\rho}_{n+1} = \tr_n \large[|\bar{2}\rangle \langle \bar{2}|_{(n,n+1)}\large]$,
 and 
\begin{equation}
\langle\xi_{\mathcal{N}}|_{n+1}= \langle \bar{2}|_{n,n+1}\langle \mathcal{N}-1|\mathcal{N}\rangle.
\end{equation} 
Extending the trace to the periodic ladder state, we get
\begin{eqnarray}
&&\rho^{(2)}_{P(n+1,n+2)}=\rho^{(2)}_{n+1,n+2} + \tr_{1..n}[|\mathcal{N}\rangle \langle\mathcal{N}|_{(2,n+1)}|\bar{2}\rangle \langle\bar{2}|_{(1,n+2)} \nonumber\\
&&\phantom{+ (|\mathcal{N}\rangle_{2,n+1}|\bar{2}\rangle}+ (|\mathcal{N}\rangle_{2,n+1}|\bar{2}\rangle_{1, n+2}\langle \mathcal{N}+2| + h. c.)]\nonumber\\
&&\quad \quad \quad = \rho^{(2)}_{n+1,n+2} + \beta^{2}_{1(n+1,n+2)} + (\beta^{2}_{2(n+1,n+2)} + h.c.).
\end{eqnarray} 
Here, 
\begin{eqnarray}
\beta^{2}_{1(n+1,n+2)} &=& \mathcal{Z}_{\mathcal{N}-1} |1\rangle \langle 1|_{(n+1)} \nonumber \\
&\otimes & \bar{\rho}_{n+2} + \mathcal{Z}_{\mathcal{N}-2} \bar{\rho}_{n+1} \otimes \bar{\rho}_{n+2}  \nonumber \\
&+& (|1\rangle \langle \xi_{\mathcal{N}-1}|_{(n+1)} \otimes \bar{\rho}_{n+2} + h. c.),
\end{eqnarray}
with 
\begin{equation}
\langle \xi_{\mathcal{N} -1}|_{n+1} = \langle \bar{2}|_{n,n+1} \langle \mathcal{N}-2|_{2..n-1}\mathcal{N}-1\rangle_{2..n}.
\end{equation} 
Further,
\begin{eqnarray}
 \beta^{2}_{2(n+1,n+2)} &=& 
|2\rangle_{n+1,n+2}\langle 1|_{n+1}\langle \xi_\mathcal{N}|_{n+2} + |2\rangle_{n+1,n+2} \nonumber\\
&&\langle \phi_\mathcal{N}|_{n+1,n+2} + 
\bar{\rho}_{n+1} \otimes |1\rangle_{n+2}\langle \xi_{\mathcal{N}-1}|_{n+2} \nonumber\\
&+& \frac{1}{\mathcal{A}} |\xi_1 \rangle _{n+1}|1 \rangle_{n+2}\langle 1|_{n+1}\langle \eta_{\mathcal{N}-1}|_{n+2},
\end{eqnarray} 
where 
\begin{eqnarray}
\langle \phi_\mathcal{N}|_{n+1,n+2}&=& \langle \bar{2}|_{1,n+2}\langle \bar{2}|_{n,n+1} \langle \mathcal{N}-2|_{2,n-1} \mathcal{N}\rangle,\\ 
\langle \eta_{\mathcal{N}-1}|_{n+2}&=& \langle \bar{2}|_{1,n+2} \langle \mathcal{N}-1|_{2,n} \mathcal{N}-1\rangle_{1,n-1}|1\rangle_{n},\,\,\,\,
\end{eqnarray}
and
$\mathcal{A}=\langle1|1\rangle =\langle\bar{1}|\bar{1}\rangle$.

The local density matrices can be calculated by using the iterations, derived below, of the complete and partial inner products defined above. 
\textcolor{red}{For $M < 6$, using (\ref{nonper})}, the normalization of the non-periodic ladder is given by 
\begin{eqnarray}
\mathcal{Z}_\mathcal{N}&=& \langle\mathcal{N}|\mathcal{N}\rangle 
\nonumber\\
&=& \mathcal{A}\mathcal{Z}_{\mathcal{N}-1} + \mathcal{B}\mathcal{Z}_{\mathcal{N}-2}+ 2\mathcal{C}\mathcal{Y}^1_{\mathcal{N}-2} + 2 \mathcal{D}\mathcal{Y}^2_{\mathcal{N}-1},
\end{eqnarray}
where $\bar{\mathcal{A}}=\langle1|\bar{1}\rangle$  and $\mathcal{B}=\langle \bar{2}|\bar{2}\rangle$. 
\(\mathcal{C}\), \(\mathcal{D}\), \(\bar{\mathcal{C}}\), and \(\bar{\mathcal{D}}\) are given by 
$\langle 1|\bar{2}\rangle = \mathcal{C}|1\rangle + \mathcal{D} |\bar{1}\rangle$ and $\langle \bar{1}|\bar{2}\rangle = 
\bar{\mathcal{C}}|1\rangle + \bar{\mathcal{D}} |\bar{1}\rangle$, where $|\bar{1}\rangle$ is a vertical rung with a periodic dimer covering between the 
topmost and lowermost sites of that rung. The other terms in the normalization can be calculated as follows: 
\begin{eqnarray}
\mathcal{Y}^1_{\mathcal{N}}&=&\langle\mathcal{N}|\mathcal{N}-1\rangle|1\rangle= \mathcal{A}\mathcal{Z}_{\mathcal{N}-1} + 
\mathcal{C}\mathcal{Y}^1_{\mathcal{N}-1} + \mathcal{D}\mathcal{Y}^2_{\mathcal{N}-1},\nonumber\\
\mathcal{Y}^2_{\mathcal{N}}&=&\langle\mathcal{N}|\mathcal{N}-1\rangle|\bar{1}\rangle= \bar{\mathcal{A}}\mathcal{Z}_{\mathcal{N}-1} 
+ \bar{\mathcal{C}}\mathcal{Y}^1_{\mathcal{N}-1} + \bar{\mathcal{D}}\mathcal{Y}^2_{\mathcal{N}-1}\nonumber. 
\end{eqnarray}
The other term in the expression for $\rho^{(2)}_{n+1,n+2}$ that is to be calculated using iterations is
\begin{eqnarray}
\langle \xi_\mathcal{N}|&=& \langle \bar{2}|_{n,n+1}\langle \mathcal{N}-1|\mathcal{N}\rangle \nonumber\\
&=& \langle 1|(\mathcal{C}A^1_\mathcal{N} + \bar{\mathcal{C}}A^2_\mathcal{N}) + \langle \bar{1}|(\mathcal{D}A^1_\mathcal{N} + \bar{\mathcal{D}}A^2_\mathcal{N}),
\end{eqnarray}
where the iteration variables are given by
\begin{eqnarray}
A_\mathcal{N}^1&=&\sum_{i=1}^\mathcal{N} \mathcal{Z}_{\mathcal{N}-i}g_{i-1},\nonumber\\
\quad A_\mathcal{N}^2&=&\sum_{i=2}^\mathcal{N} \mathcal{Z}_{\mathcal{N}-i}h_{i-1},
\end{eqnarray}
with 
\begin{eqnarray}
g_{i+1}&=&\mathcal{C}g_i+\bar{\mathcal{C}}h_i, \nonumber\\ 
h_{i+1}&=&\mathcal{D}g_i+\bar{\mathcal{D}}h_i.
\end{eqnarray}
Here, $g_0=1$, and  $h_0=0$.

The extra iterative terms in the periodic two-rung density matrix, $\rho^{(2)}_{P(n+1,n+2)}$, can be expressed as
\begin{eqnarray}
\langle \phi_\mathcal{N}|_{n+1,n+2}&=& \langle \bar{2}|_{1,n+2}\langle \bar{2}|_{n,n+1} \langle \mathcal{N}-2|_{2,n-1} \mathcal{N}\rangle \nonumber\\
&=&\langle \bar{2}|_{1,n+2}\langle \bar{2}|_{n,n+1} ( \mathcal{X}^\mathcal{N}_1 |1\rangle_{1}|1\rangle_{n} \nonumber\\
&+& \mathcal{X}^\mathcal{N}_2 |\bar{1}\rangle_{1}|1\rangle_{n}   +\mathcal{X}^\mathcal{N}_3 |1\rangle_{1}|\bar{1}\rangle_{n} + \mathcal{X}^\mathcal{N}_4 |\bar{1}\rangle_{1}|\bar{1}\rangle_{n}\nonumber\\
&+& \langle\bar{2}|_{n-2,n-1}...\langle\bar{2}_{23}|\bar{2}\rangle_{12}...|\bar{2}\rangle_{n-1,n}).
\end{eqnarray}
Here, the iterative variables, $\mathcal{X}^N_i$, are defined by using \(A_\mathcal{N}^1\) and \(A_\mathcal{N}^2\) as
\begin{eqnarray}
\label{Xn}
\mathcal{X}^\mathcal{N}_1&=&\sum g_{2i}(A^1_{\mathcal{N}-1-2i} + \mathcal{C}A^1_{\mathcal{N}-2-2i}) + \bar{g}_{2i}\mathcal{D}A^1_{\mathcal{N}-2-2i}, \nonumber\\
\mathcal{X}^\mathcal{N}_2&=&\sum g_{2i}(A^2_{\mathcal{N}-1-2i} + \mathcal{C}A^2_{\mathcal{N}-2-2i}) + \bar{g}_{2i}\mathcal{D}A^2_{\mathcal{N}-2-2i},\nonumber\\
\mathcal{X}^\mathcal{N}_3&=&\sum h_{2i}(A^1_{\mathcal{N}-1-2i} + \mathcal{C}A^1_{\mathcal{N}-2-2i}) + \bar{h}_{2i}\mathcal{D}A^1_{\mathcal{N}-2-2i}, \nonumber\\
\mathcal{X}^\mathcal{N}_4&=&\sum h_{2i}(A^2_{\mathcal{N}-1-2i} + \mathcal{C}A^2_{\mathcal{N}-2-2i}) + \bar{h}_{2i}\mathcal{D}A^2_{\mathcal{N}-2-2i},\nonumber\\
\end{eqnarray}
where the summation is from $i$ = 0 to $\mathcal{N}$. The variables \(\bar{g}_i\) and \(\bar{h}_i\) mimic 
the same relations as $g_i, h_i$, with initial conditions $\bar{g}_0=0$, $\bar{h}_0=1$. Similarly, we can derive 
\begin{eqnarray}
\langle \eta_{\mathcal{N}-1}|_{n+2}&=& \langle \bar{2}|_{1,n+2} \langle \mathcal{N}-1|_{2,n} \mathcal{N}-1\rangle_{1,n-1}|1\rangle_{n} \nonumber\\
&=& (\mathcal{A}A^1_{\mathcal{N}-1} + \bar{\mathcal{A}}A^2_{\mathcal{N}-1})(\mathcal{C}\langle1|+\mathcal{D}\langle\bar{1}|)+\langle\bar{2}|_{n+2,1}\nonumber\\
&&( \mathcal{X}^{\mathcal{N}-1}_1\langle1|_2\langle1|_n + \mathcal{X}^{\mathcal{N}-1}_2\langle\bar{1}|_2\langle1|_n + \mathcal{X}^{\mathcal{N}-1}_3\nonumber\\
&&\langle1|_2\langle\bar{1}|_n + \mathcal{X}^{\mathcal{N}-1}_4\langle\bar{1}|_2\langle\bar{1}|_n)|\bar{2}\rangle_{1,2}|1\rangle_n,
\end{eqnarray}
where the iterative variables can be derived using (\ref{Xn}).
 These iterations can be used to obtain the partial density matrices of a periodic even-legged ladder, 
and hence its bipartite as well as genuine multipartite entanglement and other single and multi-site physical quantities, provided 
 the values of $\mathcal{A}$, $\bar{\mathcal{A}}$, $\mathcal{B}$, $\mathcal{C}$, 
$\bar{\mathcal{C}}$, $\mathcal{D}$, $\bar{\mathcal{D}}$, $\mathcal{Z}_1$, $\mathcal{Y}^1_1$, and $\mathcal{Y}^2_1$ are exactly calculated. 
The size of the reduced density matrices depend  on the type of the ladder considered. For example,  
a two-rung reduced density matrix for an $M$-legged ladder is a $\tau=2M$-spin matrix. 
Other reduced density matrices of spins smaller than \(\tau\), required e.g. for calculating the GGM, 
can be obtained from the \(\tau\)-spin matrix by partial tracings.

\textcolor{red}{For $M\geq6$, the iterations involve algebra that is slightly more complicated. $\langle 1|\bar{2}\rangle =\sum_{i=1}^k \alpha^1_i |\alpha_i \rangle$ 
where $|\alpha_1\rangle=|1\rangle$ and $|\alpha_2\rangle=|\bar{1}\rangle$. $|\alpha_k\rangle$ ($k\neq$ 1, 2) are other singlet combinations of an $M$ legged single vertical rung. For $M=$ 4 (in the previous derivation), $\alpha_k^1$ ($k\neq$ 1, 2) = 0. $\alpha_1^1$ and $\alpha_2^1$ are $\mathcal{C}$ and $\mathcal{D}$ respectively. Hence, for $M\geq6$, we have the following relation;
$
_{n}\langle \alpha_k|\bar{2}\rangle_{n,n+1}=(-1)^{n-1}\sum_j \alpha_j^k |\alpha_j\rangle.
$
The normalization terms then work out to be, 
\begin{eqnarray}
\mathcal{Z}_\mathcal{N}&=& \langle\mathcal{N}|\mathcal{N}\rangle_n \nonumber\\
&=& \mathcal{A}\mathcal{Z}_{\mathcal{N}-1}+\mathcal{B}\mathcal{Z}_{\mathcal{N}-2}+ 2 (-1)^{n-1}\sum_j \alpha^1_j \mathcal{Y}^j_{\mathcal{N}-1}\\
\mathcal{Y}_\mathcal{N}^j&=& _n\langle \alpha_j|(_{1,n-1}\langle\mathcal{N}-1|\mathcal{N}\rangle_{1,n})\nonumber\\
&=& \mathcal{A}_{j1}\mathcal{Z}_{\mathcal{N}-1}+ (-1)^{n-1} \sum_k \alpha^j_k \mathcal{Y}^k_{\mathcal{N}-1}
\end{eqnarray}
where $\mathcal{A}_{ij}= \langle\alpha_i|\alpha_j\rangle$. The rest of the iterations can be derived using these terms in a similar fashion to the derivation done for $M<6$. The terms $\alpha_i^j$($i,j=$ 1 to $k$), $\mathcal{A}$, $\mathcal{B}$ and $\mathcal{A}_{ij}$ need to be exactly calculated. 
}

\textcolor{blue}{One of the main motivations for the present work is that spin ladders of the type discussed in this paper can be implemented in the laboratories. It is 
therefore important to find out whether the obtained effects are resilient to noise effects, like small admixtures of higher multiplets or of 
singlets that are not of nearest neighbors. Since the genuine multiparty entanglement that we use in the paper is a continuous function of the state parameters,
such small admixtures of noise 
will not substantially alter the amount of the measure present in the multiparty state. This feature is valid independent of whether we are considering 
even- or odd-legged ladders.}

\section{DMRM: The Odd Ladder}
\label{odd}
The periodic recursion for the odd ladder is rather different from that of the even ladder. This is due to the fact that there exists no analogous state for $|1\rangle$ 
in the odd ladder. The non-periodic $|\mathcal{N}\rangle$ can be written as a series of $\mathcal{N}=2$ odd-legged ladders. 
$|\mathcal{N}\rangle=|2 \rangle_{1,2} |2 \rangle_{3,4}...|2\rangle_{n-1,n}$. 
The periodic recursion for the odd $M$-legged $\mathcal{N}$-spin-per-chain ladder is then given by 
\begin{eqnarray}
\label{operiodic}
|\mathcal{N}+2\rangle_P = |\mathcal{N}\rangle_{1,n} |2\rangle_{n+1,n+2} + |\mathcal{N}\rangle_{2,n+1} |2\rangle_{n+2,1}
\end{eqnarray}
where $|2\rangle_{n+1,n+2}$ is a two-rung odd-legged ladder at rungs $n+1,n+2$. \textcolor{red}{We observe that there is no repetition of terms in $|\mathcal{N}+2\rangle_P$ and hence no subtraction of states is required, as is needed for the even ladder. The repetition is avoided by using the periodic condition in the initial recursion. This is possible due to the absence of $|1\rangle$ state in the odd ladder}.

The density matrix can now be calculated
as before, using 
(\ref{operiodic}): 
$\rho^{\mathcal{N}+2}_P = |\mathcal{N}+2\rangle \langle \mathcal{N}+2|_P$. The reduced density matrices can be obtained by tracing out the
requisite number of spins. 
In particular, for 
obtaining 
a two-rung reduced density matrix, we  trace out the rungs ranging from 1 to \(n\): 
\begin{eqnarray}
&&\rho^{(2)}_{P(n+1,n+2)}= \tr_{1..n}[ |\mathcal{N}+2\rangle \langle \mathcal{N}+2|_P]\nonumber\\
&& = \tr_{1..n}[|\mathcal{N}\rangle \langle \mathcal{N}|_{1,n}|2 \rangle \langle 2|_{n+1,n+2} 
+ |\mathcal{N}\rangle \langle \mathcal{N}|_{2,n+1}|2 \rangle \langle 2|_{n+2,1} \nonumber\\
&& + (|\mathcal{N}\rangle_{1,n}|2\rangle_{n+1,n+2}\langle \mathcal{N}|_{2,n+1}\langle 2|_{1,n+2} + h. c.)]. \nonumber\\
\end{eqnarray}
After simplification, the above equation reads
\begin{eqnarray}
\label{p}
\rho^{(2)}_{P(n+1,n+2)} &=&\mathcal{Z}_\mathcal{N} |2\rangle \langle 2|_{(n+1,n+2)} + \mathcal{Z}_{\mathcal{N}-2} \bar{\rho}_{n+1} \otimes \bar{\rho}_{n+2}\nonumber\\
& +& (|2\rangle_{n+1,n+2}\langle \Omega_\mathcal{N}|_{n+1,n+2} + h. c.),
\end{eqnarray}
where 
\begin{equation}
\langle \Omega_\mathcal{N}|_{n+1,n+2}=\langle 2|_{1,n+2}\langle \mathcal{N}|_{2,n+1}|\mathcal{N}\rangle_{1,n},
\end{equation}
 and the normalization is 
\(\mathcal{Z_N} = \langle \mathcal{N}| \mathcal{N}\rangle
= \mathcal{Z}_2^{N/2}\).
Now, for a $M$-legged ladder, $\mathcal{Z}_2$ corresponds to a two-legged $M$-site-per-chain ladder, and can be calculated from the previous section. 
The recursion for $\langle \Omega_\mathcal{N}|$ is 
\(\langle \Omega_\mathcal{N}|_{n+1,n+2} = \langle 2|_{1,n+2}\langle \mathcal{N}|_{2,n+1}|\mathcal{N}\rangle_{1,n} 
= \langle 2|_{1,n+2} \langle 2|_{2,3}...\langle 2|_{n,n+1}|2 \rangle_{1,2} 
|2 \rangle_{3,4}...|2\rangle_{n-1,n}\).            
Hence the computation involves writing an algorithm to  iterate the step $\langle 2|_{i,i+3}\langle 2|_{i+1,i+2}|2 \rangle_{i,i+1}$.
Once the reduced density matrices are obtained by  the iterative method, we can again use them to obtain the different single- and multi-site physical 
quantities of the system.

\section{Even versus Odd}
\label{sec-evo}
To illustrate the effectiveness of the DMRM,
we apply it to obtain a multisite entanglement of 
%
multi-legged ladders with both even and odd number of legs.
In particular, we consider two- and four-legged ladders among even ladders, and three- and five-legged ladders among odd ones.
%
The iterative variables 
can be evaluated by using their explicitly calculated initial 
set of values. The iterations provide the reduced density matrices of the system, which are thereafter utilized to obtain the GGM.  

Specifically, 
for $M=2$, the relevant initial 
parameters are 
\begin{eqnarray}
\mathcal{Z}_0=1, \quad \mathcal{Z}_1=2, \nonumber \\
\mathcal{A}=2, \quad \bar{\mathcal{A}}=2, \nonumber \\
\mathcal{C}=1, \quad \mathcal{D}=0, \quad \bar{\mathcal{C}}=0, \quad \bar{\mathcal{D}}=0, \nonumber \\
\mathcal{Y}_1^1=2, \quad \mathcal{Y}^2_1=0. 
\end{eqnarray}
%
For $M=4$, the initial parameters are 
\begin{eqnarray}
 \mathcal{Z}_0=1, \quad \mathcal{Z}_1=4, \nonumber \\
\mathcal{A}=4, \quad \bar{\mathcal{A}}=2, \nonumber \\
\mathcal{C}=5, \quad \mathcal{D}=1, \quad \bar{\mathcal{C}}=2, \quad \bar{\mathcal{D}}=3, \nonumber \\
\mathcal{Y}_1^1=4, \quad \mathcal{Y}^2_1=2. 
\end{eqnarray}
A similar analysis can also be done for higher even-legged ladders.
\begin{figure}[htb]
\label{fig:2}
\begin{center}
\epsfig{figure= 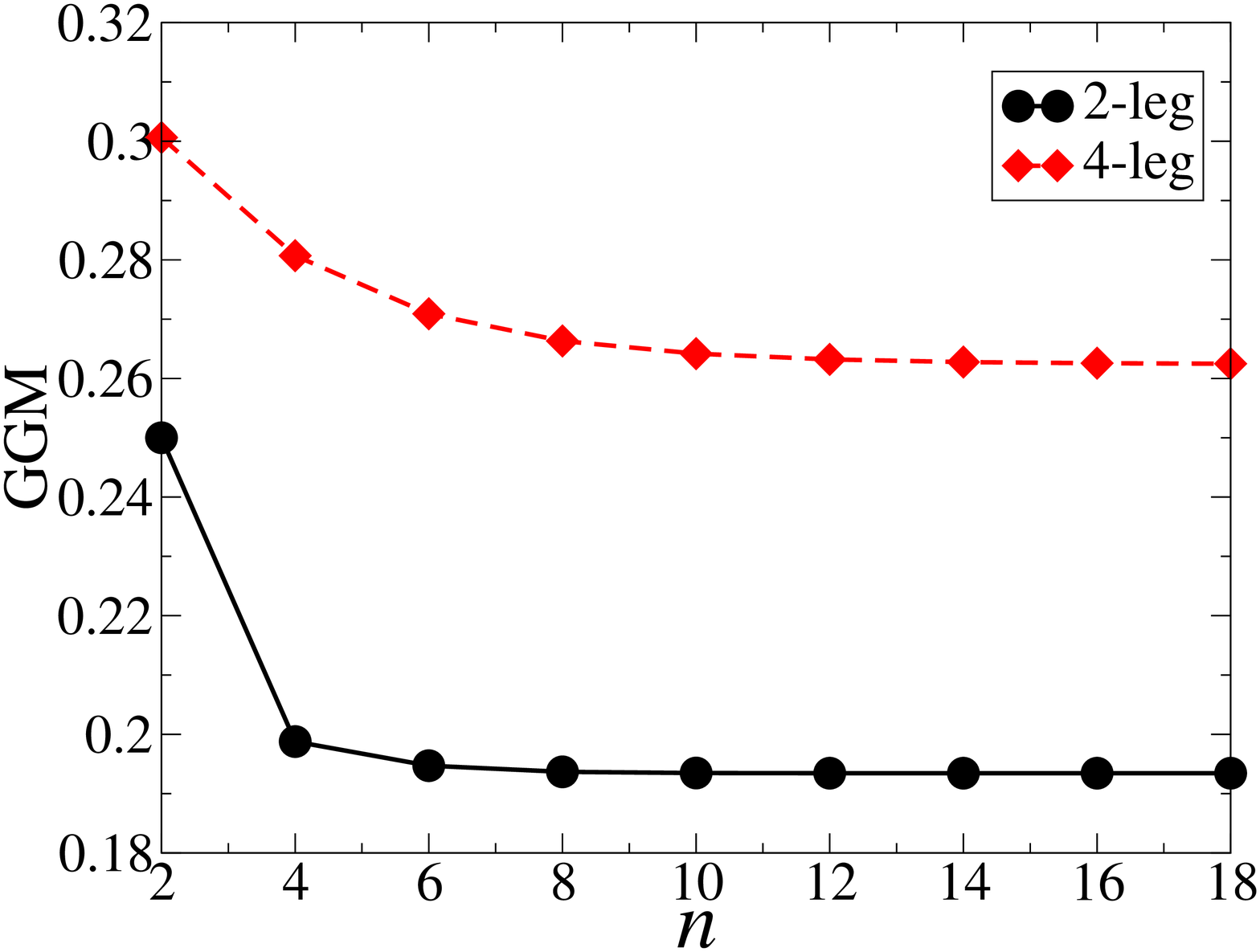, height=.2\textheight, width=0.45\textwidth}
\caption{
Genuine multisite entanglement decreases with system-size for even ladders. We perform the iterations for \(M=2\) and \(4\). The 
iterations are carried out until 18 rungs, i.e. 36 and 72 sites respectively for \(M=2\) and 4. 
However in both the cases, the GGM converges much before those sizes. 
The vertical axis represents the GGM, while the horizontal one represents the number of rungs. Both axes are dimensionless.
}
\end{center}
\end{figure}
In the case of odd ladders, the initial parameter required in the recursion 
for $M=3$ is $\mathcal{Z}_2=44$, and for $M=5$ is $\mathcal{Z}_2=804$. 

We are now ready to compare the multisite entanglements for the even- and odd-legged ladders. The GGMs obtained by the iterative methods 
clearly capture the characteristic complementary nature of even and odd ladders. We show that the GGM decreases with the increase of system-size in the case of 
even ladders (Fig. 2). The opposite is true for odd ladders -- the GGM increases with system-size (Fig. 3). 
\begin{figure}[htb]
\label{fig:3}
\begin{center}
\epsfig{figure= 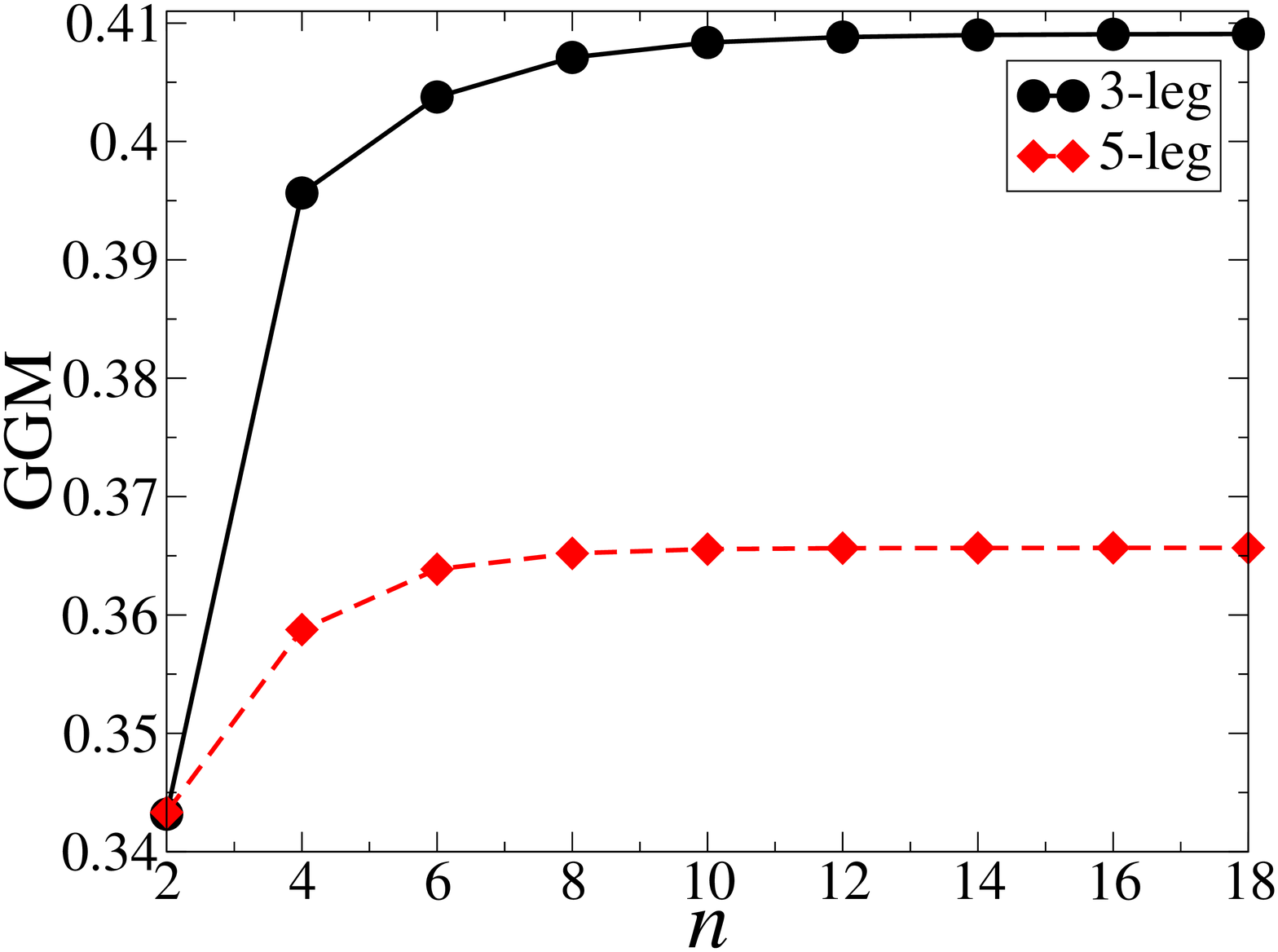, height=.2\textheight, width=0.45\textwidth}
\caption{ 
Genuine multisite entanglement increases with system-size for odd ladders. We perform the iterations for \(M=3\) and \(5\). The 
iterations are until 18 rungs, i.e. 54 and 90 sites respectively for \(M=3\) and 5. 
Again in both the cases, the GGM converges much before 18 rungs. 
The vertical axis represents the GGM, while the horizontal one represents the number of rungs. Both axes are dimensionless.
%
%
}
\end{center}
\end{figure}

Although the comparison is made by taking \(M=3\) and 5 among 
odd ladders, we have actually performed the computations also for \(M=1\), which again shows the characteristic increasing GGM of odd ladders. In all the cases, 
the GGM is calculated by considering reduced density matrices upto \(2M\) spins: numerical exact diagonalizations corroborate that considering upto 4 spins is already enough.

We have performed the iterative algorithms upto 18 rungs in all the cases (which e.g. is equivalent to 72 spins for the four-legged and 90 spins 
for the five-legged ladder). As seen in the Figs. 2 and 3, the GGM has already converged much before the maximum number of rungs that we have considered. 

\textcolor{red}{The entanglement properties of the quantum spin ladder states can be generalized and scaled to study other important aspects of quantum spin systems. The two-site reduced density matrix of superposed dimer-covered state is a Werner state \cite{eita-Werner} and the bipartite entanglement of the state is a monotone of the Werner parameter. The Werner parameter can also be used to study the ground state energy and spin correlation length (see e.g. \cite{Fan, del, del2} and references therein). 
There is also a  correspondence between the entropic properties of spin ladder states with entanglement \cite{Kallin}.}


\section{Conclusion} 
\label{sec-conclu}
We have introduced an analytical iterative technique, the density matrix recursion method, 
which can be efficiently used to obtain arbitrary reduced density matrices of the states of spin-1/2 \textcolor{red}{quantum spin ladders with an 
arbitrary number of legs, formed by the superposition of dimer coverings.}
This technique immediately allows us to obtain single- and multi-site physical properties of the system. 
In particular, we use the method to obtain the scaling of genuine multisite entanglement in the states of both odd- and even-legged ladders. 
We find that the genuine multisite entanglement can capture the disparity between the even and odd ladder states. 
\textcolor{red}{These entanglement properties may prove insightful to researchers interested in studying other physical aspects of spin systems.}
\textcolor{blue}{It would be interesting to find the extrapolation of the results obtained to the case of broad rungs.}

\textcolor{red}{The behavior of multipartite entanglement of such 
systems has created a lot of interest due to 
recent experimental developments. Simulating large qubit systems, using optical superlattices \cite{OL} and interacting photon states \cite{WP}, to generate dimer-covered superposed states in the laboratory points to the future applications of multi-partite entanglement properties of these states. These multipartite entangled states can potentially find applications in building cluster states for large scale quantum computation \cite{com} and in quantum metrology \cite{met}.}

\textcolor{red}{Our analytical method enables the investigation of the bipartite and multipartite entanglement properties in these superposed dimer-covered systems with relative control and ease even for systems containing a considerable number of quantum spins. 
The results obtained may prove useful in predicting the prospects of the entanglement properties of experimentally generated states for future applications.}
%

\ack
HSD is supported by the University Grants Commission (UGC), India under the UGC-Senior
Research Fellowship scheme. HSD thanks the Harish-Chandra Research Institute for hospitality
and support during visits. Computations were performed at the cluster computing facility at HRI
and at the UGC-DSA computing facility at Jawaharlal Nehru University.

\end{document}